\documentclass[runningheads]{llncs}
\usepackage[T1]{fontenc}
\usepackage{subfigure}
\usepackage{hyperref}
\usepackage{todonotes}
\usepackage{graphicx}
\begin{document}
\title{Lightweight Deep Models for Dermatological Disease Detection: A Study on Instance Selection and Channel Optimization}
%
%
\author{Ian Mateos Gonzalez \and Estefani Jaramilla Nava \and Abraham Sánchez Morales \and Jesús García Ramírez \and Ricardo Ramos Aguilar}
%
%
\institute{No given}
\institute{Unidad Profesional Interdisciplinaria de Ingeniería Campus Tlaxcala, Guillermo Valle 11, Centro, C.P. 90000 Tlaxcala de Xicohténcatl, Tlaxcala.\\
\email{\{imateosg2200, ejaramillon2200, amoraless2201\}@alumno.ipn.mx}, \{jegarciara, rramosa\}@ipn.mx}
\maketitle              
\begin{abstract}
The identification of dermatological disease is an important problem in Mexico according with different studies. Several works in literature use the datasets of different repositories without applying a study of the data behavior, especially in medical images domain. In this work, we propose a methodology to preprocess dermaMNIST dataset in order to improve its quality for the classification stage, where we use lightweight convolutional neural networks. In our results, we reduce the number of instances for the neural network training obtaining a similar performance of models as ResNet.

\end{abstract}
\section{Introduction}

Dermatological diseases affect a significant portion of the population worldwide. According to Universidad Nacional Autónoma de México the skin cancer was the second more frequent in the country~\footnote{\href{https://www.fundacionunam.org.mx/unam-al-dia/cancer-de-piel-segundo-mas-frecuente-en-mexico/}{Reported in "UNAM al día"}}. Given the prevalence of these diseases and the need for accurate identification, we propose a deep learning-based strategy for their detection and classification.

Some studies in the literature implement deep learning models directly, often without conducting a thorough analysis of the datasets used. Some of them propose an statistical analysis of the datasets~\cite{mohammed2024statistical,jones2011multiple}, and only one study the quality of images in DermaMNIST dataset. In this work, we utilize DermaMNIST, a dataset from the MedMNIST archive~\cite{medmnistv2} that includes various dermatological diseases as a classification problem. Unlike other studies, we propose performing an in-depth analysis of the dataset, applying instance selection techniques, and employing data augmentation to enhance model performance.

Additionally, we propose to implement lightweight convolutional neural networks. We hypothesize that we can use these models achieving comparable performance to classical models such as VGG~\cite{simonyan2014very} or ResNet~\cite{he2016deep}. Furthermore, we consider that utilizing less complex models can outperform or obtain similar performance compared with highly complex models. The code is available in a Github repository~\footnote{Space for the Github repository}

Our results demonstrate comparable performance to more complex models such as ResNet-18 and ResNet-50~\cite{he2016deep}. Additionally, we significantly reduce the computational cost of model training by using fewer dataset instances. The proposed analysis help us to implement our solution in a conventional computer without the need of specialized hardware as a GPUs.

This paper is organized as follows: Section~\ref{sec:methodology} describes the proposed methodology, Section~\ref{sec:ds_analysis} presents the analysis of the DermaMNIST dataset, and Section~\ref{sec:experiments} discusses the experimental results of the classification stage. Finally, Section~\ref{sec:conclusions} presents the conclusions and future work.

\section{Proposed Methodology}
\label{sec:methodology}

The methodology proposed for the classification of skin diseases is divided into two phases: (1) Data analysis, which involves understanding the composition of images in RGB format, applying statistical methods for data visualization and instance selection, allowing for reduce the necessary instances for training a model; 
(2) Experimental method, where tests were conducted with different configurations of convolutional neural networks, modifying various activation functions, aiming to find a balance between complexity and performance, which has a significant impact on the results, as demonstrated in the work developed by Rodriguez, highlighting the importance of parameter variation for a similar application~\cite{Rodriguez2021}.

\begin{figure}[htbp]
    \centering
    \includegraphics[height=2.7cm]{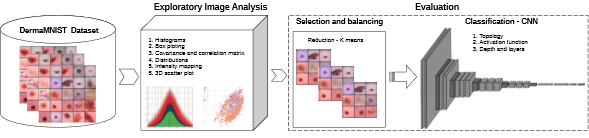}
    \caption{Distribution of the classes in DermaMNIST dataset. The acronyms corresponds to the diseases in the dataset: Melanocytic Nevi (MN), Melanoma (ME), Benign Keratosis-like lesions (BK), Basal Cell Carcinoma (BC), Actinic Keratoses (AL), Vascular Lesions (VL), Dermatofibroma (DF)}
    \label{fig:methodology}
\end{figure}

Figure~\ref{fig:methodology} shows the workflow of the methodology, highlighting two main phases: data analysis, which focuses on detecting patterns and selecting specific 
instances for 
model training. This stage reduces the computational workload required by deep learning models and improves data representation \cite{coleman2020selectionproxyefficientdata}.
This analysis stage adds importance to the used data~\cite{JEBB2017265}. Second, the evaluation stage involves train deep models, in this work we test different activation functions over the same architecture.
 
\subsection{Exploratory image analysis} \label{sec:eda}


As shown in Figure \ref{fig:methodology}, the initial stage involved data analysis to gain a general perspective of the dataset. The class imbalance was identified using histogram visualizations and the imbalance index. Additionally, a univariate analysis was performed using box plots, based on the mean intensity of each color channel per class. Subsequently, a bivariate visual analysis was conducted using a scatter matrix and a correlation matrix with the mean color channels 
of the images to observe dependent relationships. 

The next step involved visualizing the RGB channels of the entire dataset per class through distributions, reflecting the distribution of color channels with slight separations. An intensity map analysis was then performed across the entire dataset, combining classes while separating by color channel, highlighting the similarity in intensity among them. Finally, a scattered visualization of the data was generated in two and three dimensions, coloring the labels to observe the distribution of classes across channels and the relationships between channels per label.

The importance of data analysis and interpretability is a crucial stage to ensure achieving specific objectives in various machine learning tasks, such as feature extraction, feature selection, dimensionality reduction, balancing, among others. In the medical field, this is particularly important since the interpretation of data or results directly impacts patients \cite{kibria2018big}. In this context, this work proposes a classification task for different diseases in the DesmaMNIST dataset, starting with an exploratory analysis of the images using graphs and statistical methods.

\subsection{Evaluation}

Based on the exploratory image analysis mentioned in subsection \ref{sec:eda}, T-SNE was applied to gain another perspective on the distribution of the images. Based on the observations and conclusions from the analysis section, data reduction was performed by applying K-means to class five. 

The histogram of the Figure~\ref{fig:histograma} shows the following data counts: 4693, 779, 769, 359, 228, 99, and 80, respectively. A clear imbalance is evident in the dataset, with the Melanocytic Nevi class (on the left) containing the majority of the data, while the Vascular Lesions and Dermatofibroma classes (on the right) are significantly underrepresented. The calculated balance index of 0.02, which is very close to zero, further confirms the substantial disparity between the classes. 

From the obtained results, ISOMAP was used to visualize the distribution between classes with feature selection. Finally, different convolutional neural network architectures were tested, evaluating models with different activation functions to achieve a balance between complexity and performance. Additionally, hyperparameters such as batch size, learning rate, and the number of epochs were tuned. These optimizations significantly improved the models accuracy and stability, demonstrating the effectiveness of the methodology in detecting skin diseases.

\section{DermaMNIST Analysis}
\label{sec:ds_analysis}

In this section, we present an analysis of the DermaMNIST dataset. We begin by examining the number of instances in each class. Next, we provide a quantitative analysis of the dataset using data visualization techniques such as t-SNE~\cite{van2008visualizing} and Isomap~\cite{tenenbaum2000global}. Finally, we conduct a correlation study between the color channels to evaluate whether one or a combination of them can be used for the classification stage.

\subsection{Instances for each class}

\begin{figure}[htbp]
    \centering
    \includegraphics[height=4.5cm]{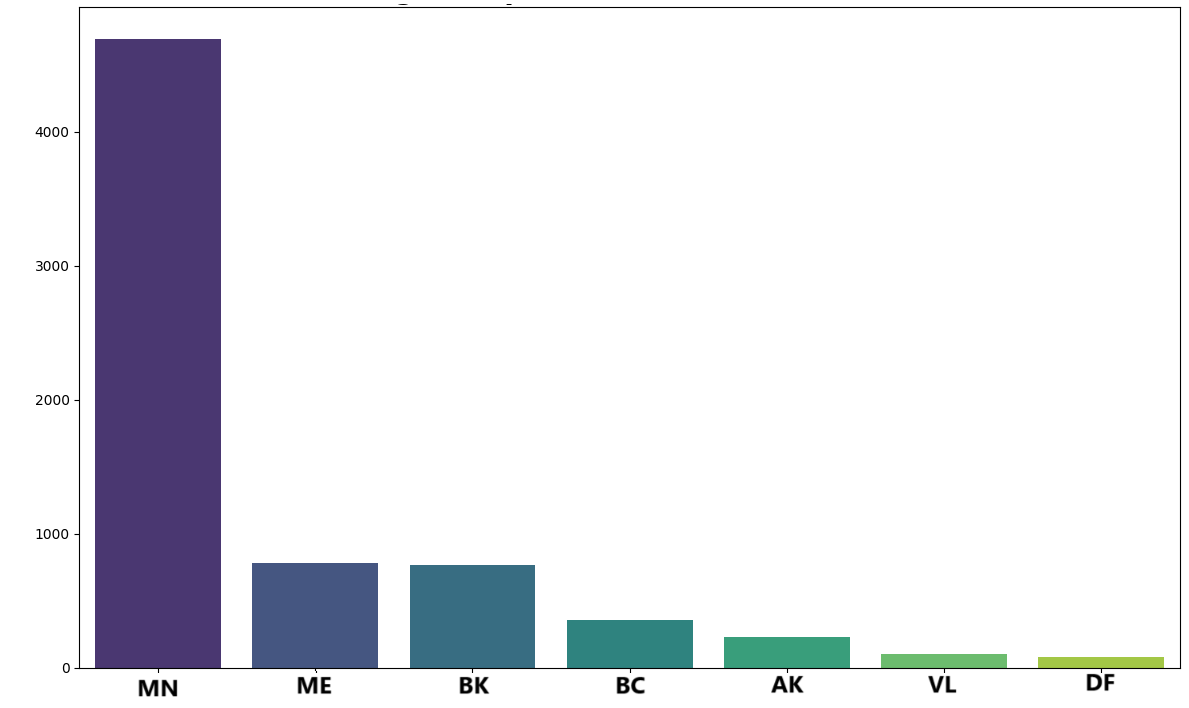}
    \caption{Distribution of the classes in DermaMNIST dataset. The acronyms corresponds to the diseases in the dataset: Melanocytic Nevi (MN), Melanoma (ME), Benign Keratosis-like lesions (BK), Basal Cell Carcinoma (BC), Actinic Keratoses (AL), Vascular Lesions (VL), Dermatofibroma (DF)}
    \label{fig:histograma}
\end{figure}

The DermaMNIST dataset consists of seven classes, each corresponding to a specific dermatological disease. The distribution of instances across these classes is shown in Fig.~\ref{fig:histograma}. It can be observed that the Melanocytic Nevi class has a significantly higher number of instances compared to others, which could pose a limitation during the classification stage. To address this issue, we propose an instance selection approach based on the k-means clustering algorithm with $k=1000$ only for this class (to apply k-means, the images are flattened and used as input features), aiming to balance the number of instances across classes. Then we use the nearest instances to the obtained centroids. For the two classes with lower number of data we will apply data augmentation.

\subsection{Data visualization}

Additionally, to visualize the dataset space, we apply two distance-preserving visualization techniques: t-SNE and Isomap. These transformations are performed both before and after the instance selection process described in the previous section. Also, the images are flattened to apply the algorithms.

\begin{figure}[t]
    \centering
     \subfigure[]{\includegraphics[width=4cm]{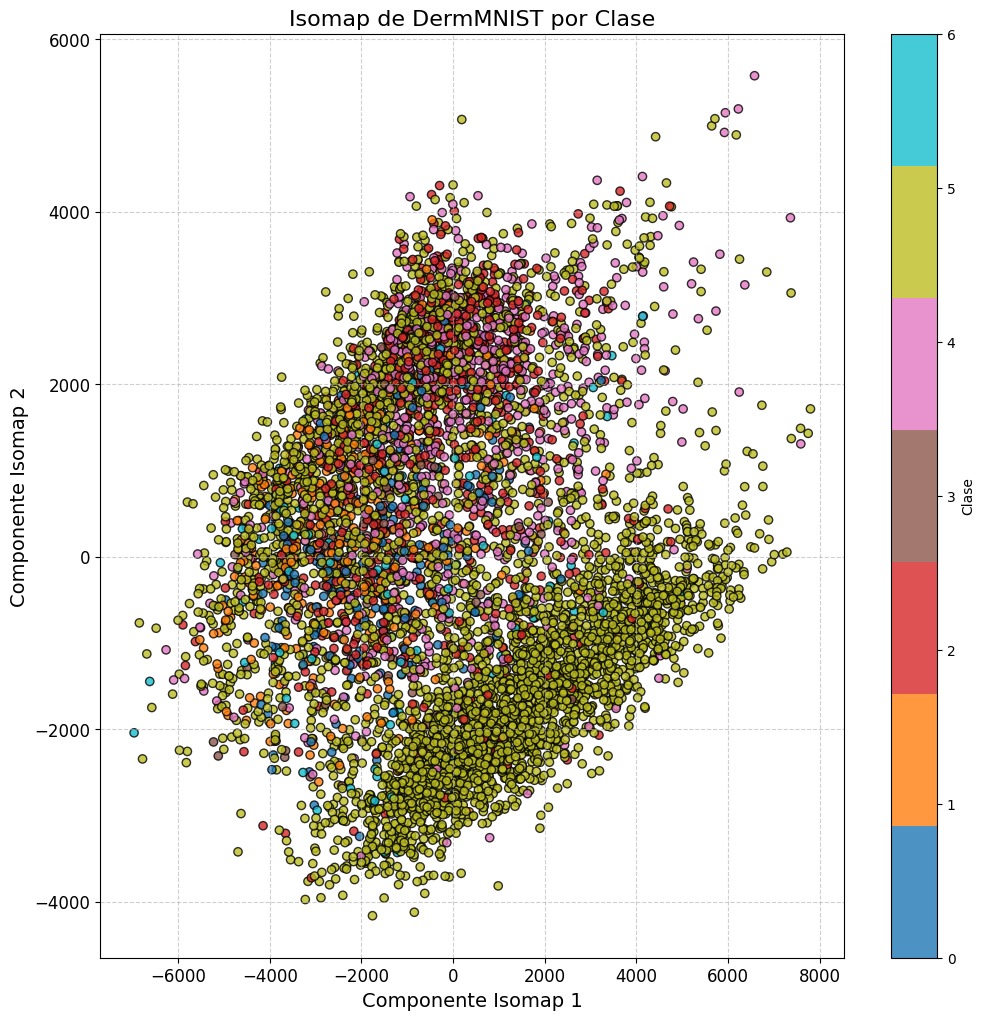}}
     \subfigure[]{\includegraphics[width=4cm]{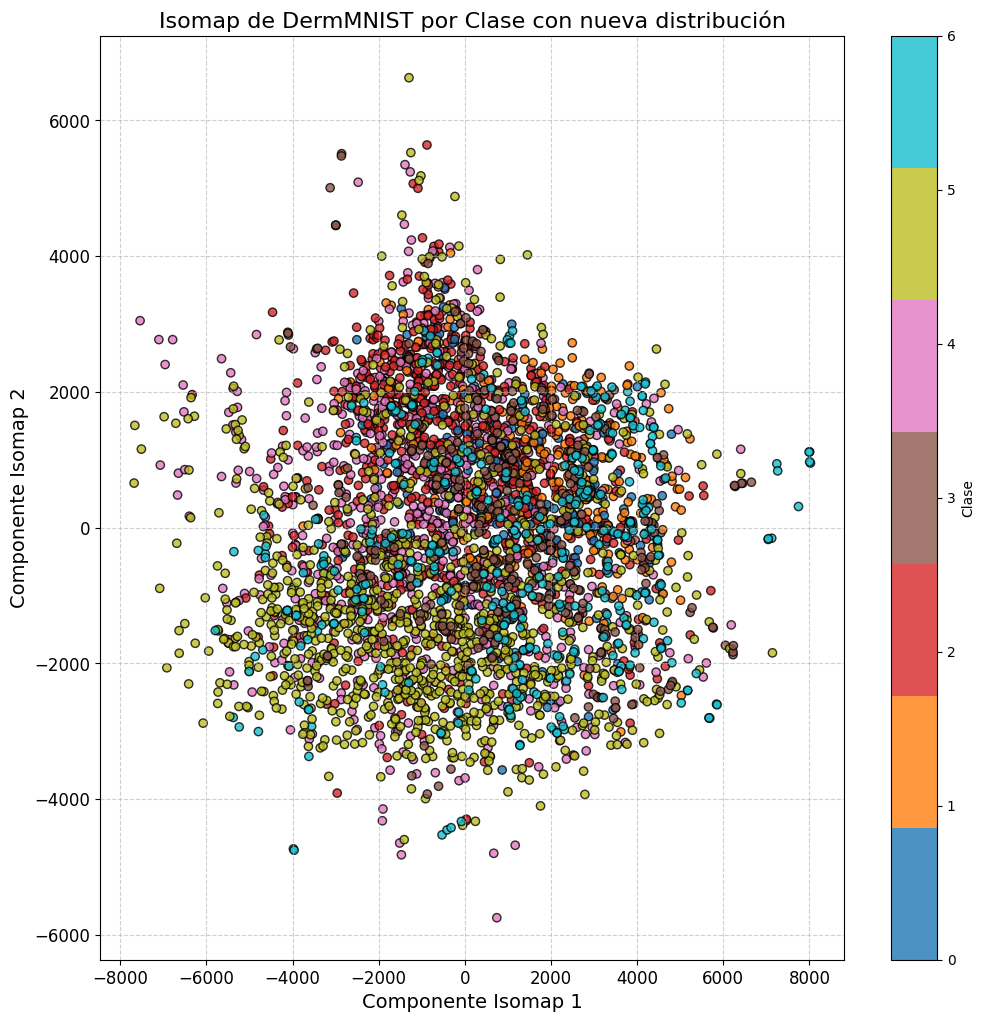}}
     \subfigure[]{\includegraphics[width=4cm] {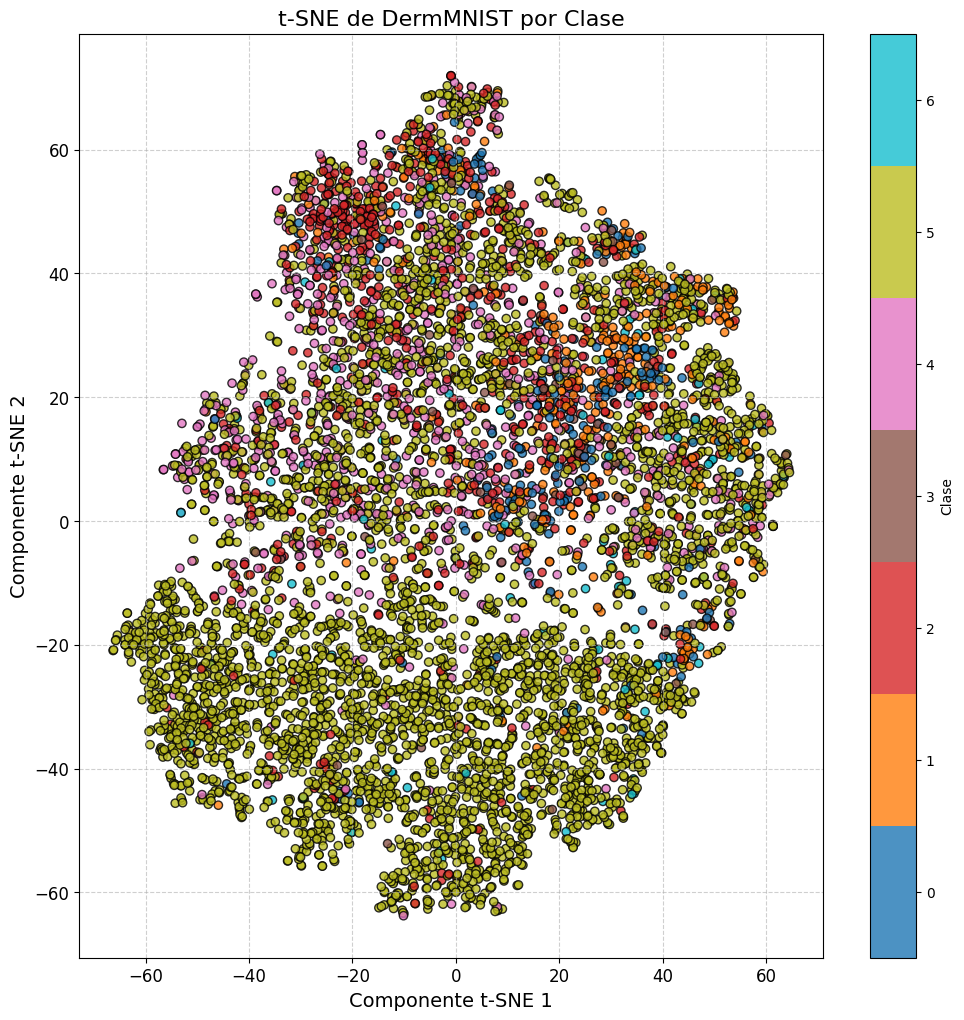}}
     \subfigure[]{\includegraphics[width=4cm]{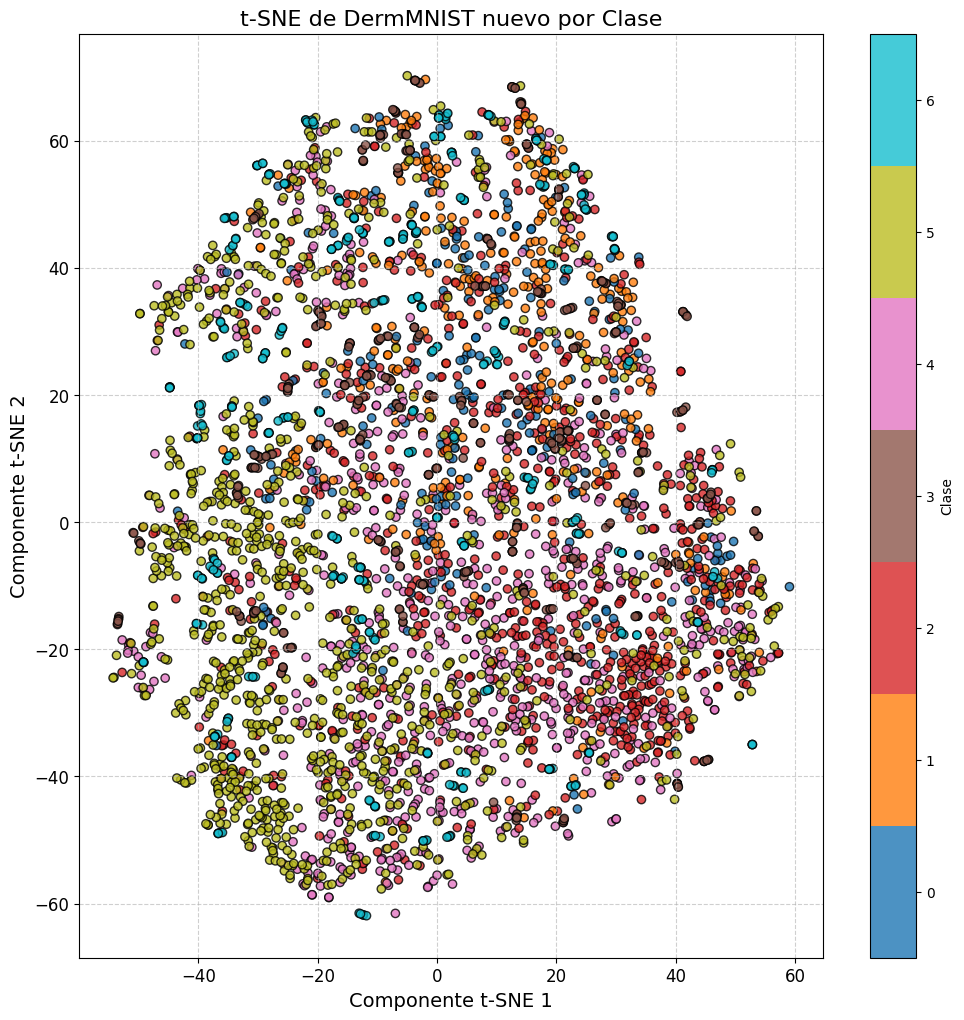}}
    \caption{Visualization of the data with t-SNE and Isomap algorithms. (a) Isomap before the transformation; (b) Isomap after the transformation; (c) t-SNE before the transformation; (d) t-SNE after the transformation.}
    \label{fig:enter-label}
\end{figure}

In both algorithms, we observe that the class with the highest number of instances is distributed across the entire dataset space. After applying the transformation, a better separation of the data is noticeable. However, the Melanocytic Nevi class remains spread throughout the entire space.

Based on the results of the visualization techniques, we conclude that the proposed preprocessing method, which involves instance selection for the class with the largest number of samples, could contribute to improving the classification of dermatological diseases.

\subsection{Channel selection}

Additionally, we analyze the correlation between the instances in the dataset. Specifically, we investigate whether there is a significant correlation among the RGB channels, with the goal of using only one or two channels during neural network experimentation to reduce computational complexity.

\begin{figure}[htbp]
    \centering
     \subfigure[]{\includegraphics[height=5cm]{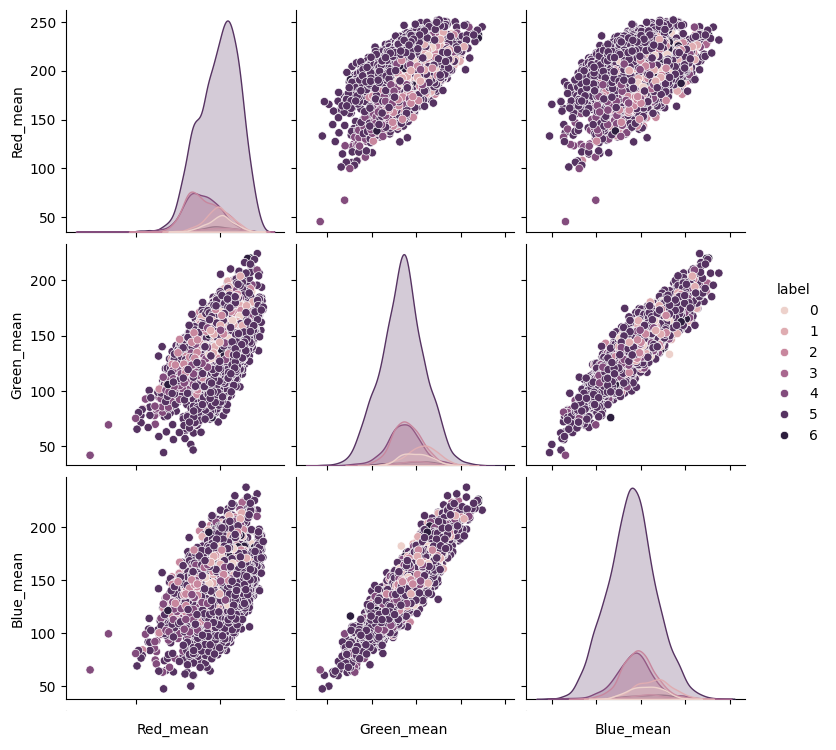}}
     \subfigure[]{\includegraphics[height=5cm]{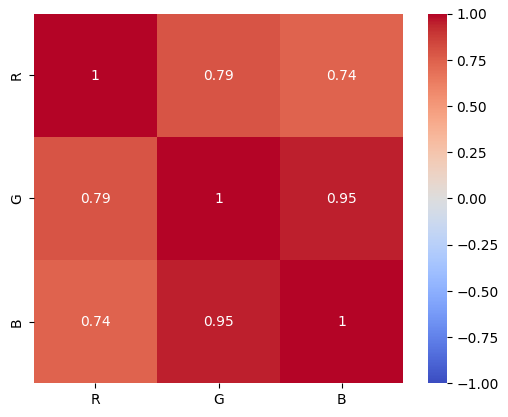}}
    \caption{Visualization of the correlation among the RGB channels. (a) Scatter plot displaying the mean values of the RGB channels, where each point represents an instance in the dataset; (b) Correlation matrix of the three channels. A strong correlation between the green and blue channels is observed.}
    \label{fig:correlation}
\end{figure}

Since the dataset consists of RGB images, we consider using a single value per channel. Specifically, we reduce each image by computing the mean of each channel and use this as a feature to compare them. Figure~\ref{fig:correlation} shows the correlation between channels using the described process. The results indicate a strong correlation between the blue and green channels, leading us to hypothesize that using only one of these two channels could achieve a similar performance to training with all three channels.

\section{Experimental Results}
\label{sec:experiments}

In this section, we present the experimental results. The objective of the experiments is to determine whether using only two channels to train a neural network can achieve a performance similar to that obtained with all three channels. 

Our experiments were conducted on a computer with an AMD Ryzen 5 processor running at 2.3 GHz and 8 GB of RAM. All experiments were performed on the CPU. After applying data preprocessing, the experiments could be executed within a reasonable runtime on this machine.

The used architecture consist of three convolutional layers with 64, 128, and 128 kernels, respectively. After each convolutional layer, we add batch normalization and max-pooling layers. In the final layers, we use two fully connected layers with 256 and 128 units. The output layer consists of 7 units with a softmax activation function. We implement three different configurations of this architecture, varying the activation function (ReLU, ELU, and GELU). The architectures contain approximately 472K parameters.

For training, we use only the selected instances along with the data augmentation techniques described in previous sections. As mentioned in the data analysis, we consider using only two channels since the green and blue channels exhibit a high correlation. The channel configurations include combining the red channel with either the blue or green channel. Additionally, we conduct experiments using all three channels for comparison. Regarding hyperparameters, we use the default settings from the Keras library, modifying only the batch size to 32 and the learning rate to 0.0001.

In our experiments, we perform five repetitions of the neural network training to compare the performance of different configurations. The mean and standard deviation of these five experiments are presented in Table~\ref{tab:results}. Our initial hypothesis was that using only two channels would be sufficient to achieve an acceptable performance. However, training with all three channels outperformed the three proposed configurations. Additionally, we observe that using only two channels leads to instability across the experiment repetitions.

\begin{table}[htbp]
    \centering
    \begin{tabular}{p{2cm}|p{2.5cm}|p{2.5cm}|p{2.5cm}}
        \hline 
              & RGB & RG & RB \\
         \hline 
         RELU & 68.83 $\pm$ 0.76 & 66.43 $\pm$ 2.20 & 65.76 $\pm$ 1.89\\
         ELU  & \textbf{69.38 $\pm$ 1.55} & 64.72 $\pm$ 1.55 & 65.41 $\pm$ 2.15\\
         GELU & 68.02 $\pm$ 1.15 & 66.86 $\pm$ 1.50 & 64.20 $\pm$ 1.50\\
         \hline 
    \end{tabular}
    \caption{Classification accuracy (\%) for different activation functions and channel configurations. The results show the mean and standard deviation over five repetitions of the experiment.}
    \label{tab:results}
\end{table}

We also compare our results with those reported in the database paper~\cite{medmnistv2}, where ResNet-18 (11.7M parameters) and ResNet-50 (25.6M parameters) were used, achieving an accuracy of 73.5\% for both models. In comparison, our best-performing model, which uses ELU activation functions, achieved a similar accuracy of 71.57\%. However, it is important to highlight that our approach utilizes a significantly more lightweight architecture (472K parameters) and a reduced number of training instances. This allows us to train the models on a low-resource computer, making our method more computationally efficient.

\section{Conclusions and Future Work}
\label{sec:conclusions}

In this study, we explored the impact of reducing the number of input channels in a convolutional neural network for dermatological image classification. Our results indicate that using only two channels, based on their correlation, does not achieve the same performance as using all three channels. The best configuration with two channels showed higher instability in the experimental repetitions, suggesting that the missing information affects the model generalization ability. 

To address the class imbalance issue in the dataset, we applied an instance selection method based on k-means clustering. Visualization techniques such as t-SNE and Isomap showed that this approach improved the separation of data points; however, the majority class (Melanocytic Nevi) remained widely distributed across the feature space. 

A lightweight convolutional neural network was designed with only 472K parameters, significantly fewer than models such as ResNet-18 (11.7M parameters) and ResNet-50 (25.6M parameters). Despite this, the best-performing model in our experiments, which used the ELU activation function, achieved an accuracy of 71.57\%, a result comparable to the 73.5\% reported for ResNet-based models~\cite{medmnistv2}. The ability to train this model on a low-resource computer without GPUs highlights its efficiency and practical applicability for real-world scenarios with computational constraints.

Regarding activation functions, we evaluated ReLU, ELU, and GELU, finding that the ELU-based configuration yielded the highest average accuracy of 69.38\%. Although different activation functions influenced performance, none of the two-channel configurations were able to outperform the three-channel approach, emphasizing the importance of retaining full color information for dermatological image classification.

For future work, we aim to implement alternative selection strategies or data augmentation techniques to improve class balance in the dataset. Additionally, we plan to explore different neural network architectures and key hyperparameters to achieve competitive performance with state-of-the-art models.

\bibliographystyle{splncs04}
\bibliography{bib}

\end{document}